\documentclass[%
 reprint,
 amsmath,amssymb,
 aps,
]{revtex4-2}

\usepackage{graphicx}% Include figure files
\usepackage{dcolumn}% Align table columns on the decimal point
\usepackage{bm}% bold math
\begin{document}

\preprint{}

\title{Information loss from dimensionality reduction in 5D-Gaussian spectral data}% Force line breaks with \\

\author{Alexej Schelle}
\affiliation{IU Internationale Hochschule, Juri-Gagarin-Ring 152, D-99084 Erfurt}
\affiliation{hema.to GmbH, Ainmiller Str. 22, 80801 Munich}

\author{Hannes Lueling}
\affiliation{hema.to GmbH, Ainmiller Str. 22, 80801 Munich}

\date{\today}% It is always \today, today,
             %  but any date may be explicitly specified

\begin{abstract}

Understanding the loss of information in spectral analytics is a crucial first step towards finding root causes for failures and uncertainties using spectral data in artificial intelligence models built from modern complex data science applications. Here, we show from an elementary Shannon entropy model analysis with quantum statistics of Gaussian distributed spectral data, that the relative loss of information from dimensionality reduction due to the projection of an initial five-dimensional dataset onto two-dimensional diagrams is less than one percent in the parameter range of small data sets with sample sizes on the order of few hundred data samples. From our analysis, we also conclude that the density and expectation value of the entropy probability distribution increases with the sample number and sample size using artificial data models derived from random sampling Monte Carlo simulation methods.     

\begin{description}
\item[Purpose]
Scholarly article on arXiv. Whitepaper on GitHub.
\end{description}
\end{abstract}

%\keywords{Suggested keywords}%Use show keys class option if keyword
                              %display desired
\maketitle

%\tableofcontents

\section{\label{sec:level2} Introduction}

Artificial intelligence models are a modern and innovative tool for quantitative predictions and environmental modeling in a very large range of physical and economic applications. Typically based on data training with high-dimensional representations of state vectors that describe the system under study and typically cannot be directly related to the actual physical processes with simple equations describing the latter, AI models are most favorably implemented in modern IT technologies and environments, such as Python Machine Learning or BigML. Interpreting and applying predictions of AI models requires a solid quantitative understanding of the underlying accuracy and numerical limits of the applied algorithms as well as the underlying mathematical assumptions and methods. Despite that artificial intelligence models have been applied so far for predictive analytics in a very large range of applications, such as medicine, economics, or weather forecasting \cite{reference-0}, a current problem based on artificial intelligence technologies, programming languages and modules, such as Python Machine Learning and Big Data, is that most artificial intelligence models are not generalizable in the sense that once trained to good accuracy based on a specific data set, the AI model cannot be applied to similar datasets with slightly different parameters without firstly training the model to the new dataset. 

One specific unresolved question is whether predictions from the training of AI models with multi-dimensional data can be compared to the manual analysis of the same data with reduced dimensionality. In this context, it is especially important to relate the accuracy of model predictions in any trial of the model realization, especially when applied to disease diagnostics, which is a realistic example of a situation, where numerical model predictions are made from high-dimensional data. In contrast, the manual analysis is performed based on two- or three-dimensional datasets. As a simple example, one may manually compare predictions of an artificial intelligence software such as hema.to describe probabilities for leukemia diagnosis in terms of multi-dimensional input vectors to the standard diagnosis made from combinations of two-dimensional representations of 
cytograms \cite{reference-1}. 

\begin{figure}[t]
\begin{center}
\includegraphics[width=6.0cm, height=4.5cm,angle=0.0]{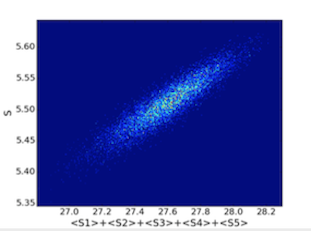}
\caption{(color online) Figure shows $10^4$ realizations of the total Shannon entropy versus the sum of conditional entropies in an artificial setup of $N=400$ particles containing 5 frequency components with corresponding photon occupation numbers $N_k$. The mean value of the distribution is around $27.6$ for the sum of conditional entropies, and $5.5$ for the total entropy, and thus the relative information loss per frequency component measured by the Shannon entropy, as defined in Eq. (\ref{entropy}) is less than one percent.}
\label{figone}
\end{center}
\end{figure}

Generally, there exists a large range of relevant software programs for generating AI model predictions based on artificial intelligence models built from spectral data, which either intrinsically make use of a dimensionality reduction in the application and analysis of the model results, or reduce the initial data to a Gaussian centered dataset using data modeling methods, such as principal component analysis. For multi-dimensional datasets, where standard accuracy and sensitivity measures defined from correlations of true-positive and true-negative outcomes only measure the accuracy of pairwise correlations between two predicted classes, a more complex measure is needed to prove that the standard accuracy and sensitivity measure (defined only pairwise on two classes) do correctly quantify the correlations between the multi-dimensional input vector which the AI model was trained on and the manual outcome analytics of an observer which compares his observations with the model predictions. In our case, the convergence of a specific mathematical quantity, the entropy calculated from a multi-dimensional probability measure for the system to satisfy certain model properties is used as a generalized measure for the accuracy of the AI model with parameters defined from an artificial multi-dimensional dataset with a given initial shape of the probability distribution. 

\begin{figure}[t]
\begin{center}
\includegraphics[width=6.0cm, height=4.5cm,angle=0.0]{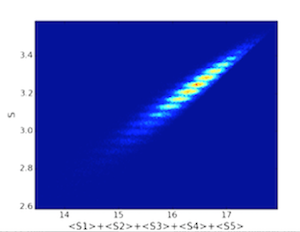}
\caption{(color online) Figure shows $10^5$ realizations of the total Shannon entropy versus the sum of conditional entropies in an artificial setup of $N=40$ particles containing 5 components with corresponding photon occupation numbers $N_k$. The structure of the entropy distribution can be better recognized as the number of sampling steps increases.}
\label{figtwo}
\end{center}
\end{figure}

Understanding the predictive outcome of an artificial intelligence model based on spectral data is often performed by analyzing the event density and intensity patterns and structures manually, to relate them to specific classifications, such as different types of disease classes \cite{reference-2}. In this type of analysis, it is crucial to understand the accuracy both of the predictive model as well as the standard analysis, to provide consistent predictions also in the case, where forecasts of artificial intelligence models and standard human analysis tend to contradict. In the sequel of the present analysis, it is therefore shown within a straightforward entropy model that the deviations and losses of information present to artificial intelligence models based on Gaussian spectral data are less than one percent from the loss of information obtained by projections of the initial multi-dimensional state vector onto reduced representations of the multi-dimensional dataset. For this purpose, in the present study, we reduce our analysis to the case of a 5-dimensional Gaussian-distributed spectral dataset, and calculate the conditional (information) Shannon entropy for the case of reduced, two-dimensional representations of the dataset, i. e. the sum of single conditional entropies versus the total correlated Shannon entropy.    

\section{\label{sec:level2} Theory}

In the present work, modeling the entropy of N-dimensional spectral data within Gaussian quantum statistics is assumed to highlight the distribution of information in the context of photon-emitting particles in a composite quantum system. As a starting point for our analysis of information loss from dimensionality reduction in such spectral representations of the multi-dimensional matrix representation in a spectral data file for data analytics, as a measure for information, we use the Shannon entropy, defined as 

\begin{equation}
S = -\sum_i p_i~{\rm log}~p_i \ , 
\label{entropy}
\end{equation}\\
where $p_i = p_i (E)$ is the probability 

\begin{equation}
p_i (E) = \prod_{k=1}^m p^{(i)}_k (E) = \mathcal{Z}^{-1}\prod_{k=1}^m{\rm e}^{-\beta E - \frac{(E-<E^{(i)}_k>)^2}{2\sigma_k^2}} \ 
\label{probability}
\end{equation}\\
for the system to occupy a quantum state with $N_k$ events of the frequency component $k$ and normalization ${\mathcal{Z}(\sigma_k^2)}$. In Eq. (\ref{probability}), 

\begin{equation}
<E^{(i)}_k> = <N^{(i)}_k>\hbar\omega_k \ , 
\label{energy}
\end{equation}\\
is the mean energy of a quantum state corresponding to the emission of $<N_k>$ photons from component $i$ with frequency $\omega_k$ on average. In the limit of large temperatures and small frequencies, Eq. (\ref{energy}) becomes a Gaussian sampling function for the number of events $N_k$ around their corresponding mean values $<N_k>$, weighted with the different frequency components $\omega_k$, which can in principle be either derived directly from spectral data analysis or declaration, providing realistic quantification of the energy distribution, as defined in Eq. (\ref{energy}).

In our present approach, for a first estimate of information loss from dimensionality reduction, we reduce our analysis to an artificial data model by sampling the distribution of entropy as defined in Eq. (\ref{entropy}) assuming artificial component frequencies in the optical range with non-trivial (average) occupations $<N_k>$ of the different frequency components. Please note that, from Eq. (\ref{probability}), it is straightforward to derive a mathematical expression for the intensity distribution of photon emission for the different components and to further implement a Monte-Carlo sampling method for the quantum energy and intensity distribution to very high accuracy. The total particle number is kept unchanged and constant during mathematical and numerical modeling \cite{reference-4}. 

To calculate the intensity distribution of the quantum distribution, i.e. photons that are emitted by the particles, we may hence define the average intensity of the light emitted from particles

\begin{equation}
<I^{(i)}_k> = \frac{<N^{(i)}_k>\hbar\omega_k}{A_i\tau_i} \ , 
\label{intensity}
\end{equation}\\ 
where $A_i$ is the effective surface of the $i^{\rm th}$ particle (defined by the effective Rayleigh surface for the different quantum particles) and $\tau_i$ is the average measurement time. The expression in Eq. (\ref{intensity}) leads to a mathematical expression of the corresponding intensity probability distribution, as a function of the component frequency and photon number occupation. 

\section{\label{sec:level3} Results and analytics}

\begin{figure}[t]
\begin{center}
\includegraphics[width=7.0cm, height=5.0cm,angle=0.0]{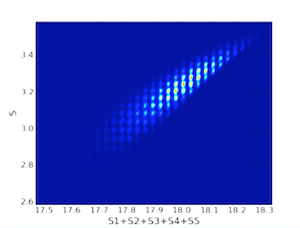}
\caption{(color online) Figure shows $10^5$ realizations of the total Shannon entropy versus the sum of uncorrelated entropies in an artificial setup of $N=40$ particles containing 5 components with corresponding photon occupation numbers $N_k$. It is observed that the relative loss of information is about five to ten times larger for (the sum of) uncorrelated Shannon 
entropies as compared to (the sum of) correlated, i.e. conditional Shannon entropies.}
\label{figthree}
\end{center}
\end{figure}

From our analysis of numerical results, we learn that relative loss of information is less than one percent starting from samples with particle numbers larger than a few hundred particles at typical optical component frequencies for the information measured by the conditional Shannon entropy, neglecting the temperature dependence for a first approximation. As shown in Fig. \ref{figone}, for a total number of $N=400$ particles, the relative loss of information as captured by the entropy measure, as defined in Eq. (\ref{entropy}), is less than one percent. The relative loss decreases for larger sample sizes and entropy density. We observe a spreading of the entropy distribution and a better connection between different local areas of the entropy distribution for a larger initial state range estimation as defined by the parameter $\delta$, i.e. the ratio of the local number uncertainty in units of the local mean photon number (expectation value) - compare Fig. \ref{figtwo}. Switching this parameter from $0.5$ to $1.0$ corresponds to assuming non-random failure modes, e. g. in the measurement process. The structure of the entropy sampling distribution is better pronounced as the number of samples increases. For an increasing number of particles, both the absolute mean value and the density of the entropy distribution increases. Numerical results are obtained from sampling Eq. (\ref{entropy}) with a Monte-Carlo sampling method, with different ranges of occupation numbers and component frequencies corresponding to the optical frequency range between $400$ and $800$ nm. Total number of particles was assumed to be in the range of $10$ to $1000$ particles. The number of frequency components was assumed to be $N = 5$. 

Notably, the quantitative scaling behavior for the case of non-conditional Shannon entropy (Fig. \ref{figthree}) behaves similarly to the shown analysis, however, the loss of information is increased to approximately ten to fifteen percent, if the components are assumed to be statistically uncorrelated. 

Finally,  from Eq. (\ref{energy}), assuming a Gaussian energy distribution of the photon energy emitted by the different particle component types, it is possible to show that the relative loss of information tends to zero when the total number of particles tends to infinity. However, even this result seems to indicate that there are almost no information losses from dimensionality reduction in spectral analytics in the limit of large particle sizes, one should keep in mind that for single event cases, the calculated ranges may still define non-negligible deviations from projection onto subsets with reduced dimensionality for certain AI model applications.

More sophisticated models and measures for information loss may in principle lead to slightly different quantitative scaling of the results.
  
\section{\label{sec:level4} Proposal and Outlook}

From the presented analysis with spectral data obtained from Markov sampling, we conclude that the levels of deviation induced by dimensionality reduction indicate that wrongly or not uniquely classified cancer cases predicted by artificial intelligence models are not likely to be induced by projecting the initial state onto reduced subspaces of the initial data space in the limit of Gaussian correlated spectral data. To understand the wrong classification of artificial intelligence models, especially in the context of disease diagnosis, which are mostly quantified on the order of zero sets, a more sophisticated understanding of the failure modes of predictions and classifications based on artificial intelligence models has to be developed.   

Financial support from hema.to GmbH for this analysis of information loss from dimensionality reduction in artificial entropy models for spectral analytics within the project grant hema.to analytics (0001) is acknowledged.

\end{document}